\documentclass[11pt]{article}
\usepackage{epsfig}
\usepackage{latexsym}

\def\bg#1{\mbox{\boldmath$#1$}}

\newcommand{\del}{\partial}
\newcommand{\beq}{\begin{eqnarray}}
\newcommand{\eeq}{\end{eqnarray}}
\newcommand{\be}{\begin{eqnarray*}}
\newcommand{\ee}{\end{eqnarray*}}
\newcommand{\bk}{{\bf k}}
\newcommand{\bp}{{\bf p}}
\newcommand{\bq}{{\bf q}}

\newcommand{\e}{\epsilon}

\newcommand{\nn}{\nonumber}
\newcommand{\ket}[1]{\mbox{$\mid\!#1\rangle$}}
\newcommand{\bra}[1]{\mbox{$\langle#1\!\mid$}}

\begin{document}

\centerline{\Large\bf {Higher Order Hadronic Energy Level Shifts in Protonium}}
\vskip 5mm
\centerline{\Large\bf {from Effective Field Theory}}
\vskip 10mm
\centerline{Xinwei Kong}
\medskip
\centerline{\it Institute of Physics, University of Oslo, N-0316 Oslo, Norway.} 
\vskip 10mm
\centerline{Finn Ravndal\footnote{Permanent address: Institute 
            of Physics, University of Oslo, N-0316 Oslo, Norway}} 
\medskip
\centerline{\it NORDITA, Blegdamsvej 17, DK-2100, Copenhagen \O .}

\bigskip
{\bf Abstract:} {\small Using effective field theory for a proton and antiproton
bound in a Coulomb potential, the shift of the ground state energy
level is calculated to arbitrary order in the scattering length. Including the next
order contact interaction, the correction due to the effective range parameter can also 
be obtained.}

\vspace{10mm}

It was first realized by Caswell and Lepage that the interactions between photons
and electrons at energies well below the electron mass can be formulated in a purely 
non-relativistic way in a theoretical framework they called NRQED\cite{CL}. 
In contrast to QED which is a renormalizable theory, NRQED is non-renormalizable. 
But used as an effective field theory\cite{eft} only for phenomena at low energies, 
it can be used to derive higher order radiative corrections in a systematic way 
without invoking covariant 
theory like the Dirac equation and the corresponding Bethe-Salpeter equation for
bound states, when the physics is inherently not relativistic\cite{KL}.
Since its formulation, NRQED has now been used by Kinoshita and Nio\cite{KN} 
to calculate higher order corrections to hyperfine splitting in muonium. Hoang, Labelle
and Zebarjad have also recently completed the order $\alpha^6$ correction of the 
hyperfine splitting of the ground state in positronium\cite{HLZ} within the same 
framework.

The effective Lagrangian is a systematic expansion in $p/M$ where $p$ is a 
characteristic momentum and $M$ is the
heavy mass in the problem, i.e. the energy above which all degrees of freedom are
integrated out. The theory is therefore only valid at momenta below $M$. In an ordinary
atom this mass is given by the electron mass, while in a hadronic atom it would be the
pion mass. Remember that the typical momentum in an atom is $\alpha m$ where $\alpha$ 
is the fine-structure constant. Each term in the expansion must obey the symmetries we
want the system to have. The expansion coefficients can be found in two ways. In an
ordinary atom where we know the underlying, fundamental theory which is QED, we can use
perturbation theory at low energies and  match 
scattering amplitudes  calculated in NRQED to those calculated in full
QED\cite{KL}. Similarly, in a hadronic atom one can
estimate the expansion coefficients by matching scattering amplitudes in the effective
theory to what one would obtain from a more fundamental theory based on exchange
of pseudoscalar and vector mesons. QCD is strongly coupled at low energies and can thus
not be used. Instead of matching to an underlying theory, one can also determine
the effective couplings  by directly comparing the obtained scattering
amplitudes with experimental data. At very low energies these are parameterized in terms
of scattering lengths and various effective ranges.

An effective field theory for non-relativistic systems of nucleons was first constructed 
by Weinberg\cite{Weinberg}. It includes systematic counting rules so that one knows which
diagrams to keep and which to discard at any order in $p/M$.
Since then, this approach has been taken up by many others\cite{Kolck}. 
In particular, Kaplan, Savage and Wise\cite{KSW_1} have studied in great
detail the nucleon-nucleon system in the $^1S_0$ state which has presented some 
theoretical problems because of the large scattering length in this particular channel. 
These problems, having to do with the removal of divergences, have now been solved by 
the same authors\cite{KSW_2} and almost simultaneously by Gegelia\cite{Gegelia}. 

We will here consider a system of proton interacting with an antiproton. The
experimental and theoretical situation has been reviewed by Batty\cite{Batty_1} and also 
related effects in other hadronic atoms\cite{Batty_2}.
At the low energies we are interested, these particles can be considered to be
without internal structure and thus be described by local Schr\"odinger fields $\psi_1$ 
and $\psi_2$. The free Lagrangian of one particles, with common mass $m_p$, is then
\beq
    {\cal L}_0(\psi) = \psi^\dagger\left(i{\del\over\del t} 
              + {1\over 2m_p}{\bg\nabla}^2\right)\psi
\eeq
In the following we will ignore the spin degrees of freedom since our results apply to
each value of the total spin quantum number of the combined system. It is described by
a Lagrangian which can be split up into three main parts,
\beq
    {\cal L} = {\cal L}_0(\psi_1) + {\cal L}_0(\psi_2) +  {\cal L}_{EM} + {\cal L}_{had}  
                                                       \label{Left}
\eeq
Here ${\cal L}_{EM}$ includes the electromagnetic interactions\cite{KL}
between the particles and ${\cal L}_{had}$ gives their hadronic interactions. When the
momenta of the particles are much smaller than the pion mass, their interactions due
to both pion and vector meson exchanges can be described by local operators\cite{Weinberg}.
Assuming in the following that the system has total angular 
momentum $L=0$, the two leading contact interactions can then be written as
\beq
      {\cal L}_{had} = - C_0 (\psi_1^\dagger\psi_1)(\psi_2^\dagger\psi_2)
      - C_2 \left[(\psi_1^\dagger{\bg\nabla}\psi_1)\cdot(\psi_2^\dagger{\bg\nabla}\psi_2)
 + h.c. \right]                                                \label{had}
\eeq
where the effective coupling $C_2$ should be of the order $C_0/M^2$. The 
second term is therefore effectively reduced by $p^2/M^2$ compared with the first. It will
be convenient to express these interactions in terms of the two-nucleon potential
\beq
    V(\bp,\bq) &=& \bra{\bp}V\ket{\bq} =  C_0 + C_2(\bp^2 + \bq^2) \nn \\
               &\equiv & V_0(\bp,\bq) + V_2(\bp,\bq)               \label{pot}
\eeq
where $\bq$ is the nucleon momentum in the initial state and $\bp$ the momentum in the
final state.

Let us now consider elastic scattering of these two hadrons due to the hadronic 
interaction. In the CM system the total energy of the two
particles is then $E = p^2/2m$ where $m$ is the reduced mass. We will first only take 
into account the contact term proportional with $C_0$ in (\ref{had}). 
\begin{figure}[htb]
 \begin{center}
  \epsfig{figure=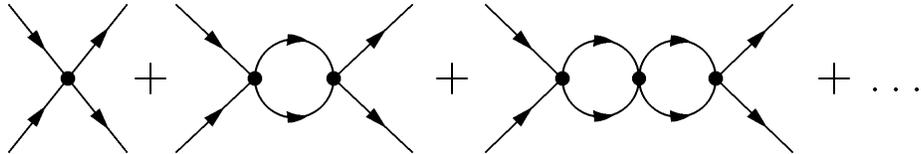,height=20mm}
 \end{center}
 \vspace{-8mm}            
 \caption{\small Chains of bubble diagrams gives the full scattering amplitude.}
 \label{fig}
\end{figure}
According to the counting rules established by Weinberg, the scattering
amplitude is given by the infinite sum of chains of bubbles as shown in Figure 1. The
sum forms a geometric series and gives the T-matrix element\cite{KSW_1}
\beq
    T_0(p) = {C_0\over 1 - C_0I(p)}                          \label{T-mat}
\eeq
Here $I(E)$ is the integral of the free two-particle propagator,
\beq
    I(p) = \int\!{d^3k\over (2\pi)^3}{2m\over \bp^2 - \bk^2 + i\e}             \label{Ip}
\eeq
It is seen to be linearly divergent in the ultraviolet. In order to regularize this
and other similar power divergences, we will use minimal subtraction at a non-zero 
momentum $\bp^2 = -\mu^2$ as suggested by Gegelia\cite{Gegelia}. This is equivalent
to PDS or power-divergence subtraction as proposed almost simultaneously by Kaplan, 
Savage and Wise\cite{KSW_2}. The result is 
\beq
    I(p) =  -{m\over 2\pi}(\mu + ip)                       \label{Ip_reg}
\eeq
When used in (\ref{T-mat}), the scattering will then depend on the {\it renormalized} 
coupling constant $C_0^R$ in this particular regularization scheme. It can be expressed 
in terms of the corresponding $S$-wave phase shift using the relations
\beq
     p\cot\delta &=& ip - {2\pi\over m}{1\over T}         \label{D1}  \\
                 &=& -{1\over a} + {1\over 2}r_0 p^2 + \ldots         \label{D2}
\eeq
when introducing the scattering length $a$ and effective range $r_0$. One then finds
\beq
    {1\over C_0^R(\mu)} = {m\over 2\pi}\left({1\over a} - \mu\right)       \label{C_0R}
\eeq
The effective range is zero in this lowest approximation.

Now consider these two hadrons of opposite electrical charges 
bound by the Coulomb force into a hadronic atom. The unperturbed $1S$ ground 
has the energy $E_0 = -\gamma^2/2m$ where $\gamma = m\alpha$ is the momentum in the
ground state with wave function $\Psi_0(r)$. It will be perturbed both by electromagnetic
and hadronic interactions. The dominant part of the energy shift $\Delta E_0$ 
will be due to the contact potential $V_0$ in (\ref{pot}). In lowest order perturbation
theory the first  contribution obviously is
\beq
     \Delta E_0^{(1)}(C_0) = \bra{\Psi_0}V_0\ket{\Psi_0} 
 = \int\!{d^3p\over (2\pi)^3}\int\!{d^3q\over (2\pi)^3}\Psi_0^*(\bp)V_0(\bp,\bq)\Psi_0(\bq)
 = C_0|\Psi_0(r=0)|^2
\eeq
where the Fourier transform of the ground state wave function is
\beq
    \Psi_0(\bp) = {8\pi^{1/2} \gamma^{5/2} \over (\bp^2 + \gamma^2)^2}   \label{Psi_0}
\eeq
Higher order contributions come from the same set of Feynman diagrams as in Figure 1. Their
contributions again form a geometric series which gives
\beq
    \Delta E_0(C_0) = |\Psi_0(0)|^2 {C_0\over 1 - C_0 I_\gamma}
\eeq
where now 
\beq
    I_\gamma = \int\!{d^3k\over (2\pi)^3}{-2m\over \bk^2 + \gamma^2}
\eeq
is the integral over  the bound two-particle propagator. It is again linearly
divergent. But after the same minimal subtraction as above, it becomes $I_\gamma^R =
-(m/2\pi)(\mu - \gamma)$. Using this result together with $|\Psi_0(0)|^2 = 
\gamma^3/\pi$ we have
\beq
    \Delta E_0(C_0) = {\gamma^3\over \pi} {C_0^R\over 1 - C_0^R I_\gamma^R}
               =  {\gamma^3\over m} {2 a\over 1 - a\gamma}
\eeq
We see that the arbitrary renormalization point $\mu$ is dropping out of the result
valid to all orders in the coupling constant as it should after renormalization.
Introducing the Bohr radius $b = 1/\gamma$ which gives the size of the atom, 
the result can be written as
\beq
    \Delta E_0(C_0) = m\alpha^2 {2a\over b - a} 
               = 2a m^2\alpha^3\left[1 + {a\over b} + \dots\right]  \label{dE1}
\eeq
since the scattering length $a \ll b$.

The leading term in this energy level shift was first derived by Deser {\it et al.}
\cite{Deser}. Since then corrections due to the electromagnetic interactions 
have been obtained by Trueman\cite{Trueman} and Kudryavtsev and Popov\cite{Popov}. These
effects can also be incorporated within the present framework and will be presented 
elsewhere\cite{KRnext}.   

We will here consider instead the effect of the higher potential $V_2$ in (\ref{pot}).
It gives a correction to the scattering amplitude which can be obtained by inserting this
interaction at every vertex in the bubble diagrams in Figure 1. One then finds 
\beq
   \Delta T(p) = {2C_2(C_0I_0 + p^2)\over [1 - C_0I(p)]^2}
\eeq
after summing all the different contributions. The divergent integral
\beq
    I_0 = -2m\int\!{d^3 k\over (2\pi)^3}
\eeq
is now zero in the regularization scheme we are using since it is independent of
external momenta. On the other  hand, $I(p)$ is the same integral as before and
takes the finite value (\ref{Ip_reg}). Inverting the full scattering amplitude
$T(p) = T_0(p) + \Delta T(p)$ and using Eqs.(\ref{D1}) and (\ref{D2}), we  find that
the renormalized coupling $C_0^R(\mu)$ is still given by (\ref{C_0R}) while 
$C_2^R(\mu)$ is related to the effective range,
\beq
     C_2^R(\mu) = {mr_0\over 8\pi} C_0^R(\mu)^2                         \label{r_0}
\eeq
as first obtained by Gegelia\cite{Gegelia} and Kaplan, Savage and Wise\cite{KSW_2}.

Now let us consider the effect of $V_2$ on the bound state problem. The lowest order 
contribution to the energy level shift becomes
\beq
     & &\Delta E_0^{(1)}(C_0,C_2) = \bra{\Psi_0}V_2\ket{\Psi_0} \nn \\
 &=& C_2(8\pi^{1/2} \gamma^{5/2})^2 \int\!{d^3p\over (2\pi)^3}\int\!{d^3q\over (2\pi)^3}
     {\bp^2 + \bq^2 \over (\bp^2 + \gamma^2)^2(\bq^2 + \gamma^2)^2}
 = 128\pi \gamma^5 C_2^R I_1 I_2^R
\eeq
Here we have introduced the convergent integral
\beq
    I_1 = \int\!{d^3p\over (2\pi)^3} {1\over (\bp^2 + \gamma^2)^2} = {1\over 8\pi\gamma}
\eeq
and the divergent integral
\beq
    I_2^R = \int\!{d^3p\over (2\pi)^3} {\bp^2\over (\bp^2 + \gamma^2)^2} 
    = {1\over 8\pi} (2\mu - 3\gamma)
\eeq
which is made finite by the same regularization method as before.
Higher order contributions from this interaction is now found by inserting it once in all 
the vertices in the bubble diagrams in Figure 1. After some resummations, we then find
\beq
     \Delta E_0(C_0,C_2) = 2{\gamma^4\over \pi}C_2^R\left[
     {2(\mu - \gamma) \over 1 - C_0^R I_\gamma^R} - 
     {\gamma\over (1 - C_0^RI_\gamma^R)^2}\right]
\eeq
Because of the coupling constant renormalization condition (\ref{r_0}), the last term
is independent of the renormalization point $\mu$. However, it appears explicitly in the 
first term so this part of the result is not invariant under the corresponding 
renormalization group\cite{KSW_2}. It is unusual to see the renormalization point appear
linearly. In renormalizable theories it always appears logarithmically since the 
ultraviolet divergences there are logarithmic. The present effective theory, on the other
hand, has power divergences in the ultraviolet which translate into explicit 
$\mu$-dependence in the corresponding beta-functions for the running coupling 
constants\cite{KSW_2}. We therefore have to accept also the explicit $\mu$-dependence 
in our result.

However, as in more conventional renormalizable theories when we are working to a finite
order in perturbation theory, we  cannot expect our result, obtained here to first order  
in  the coupling  constant $C_2$, to be invariant under the renormalization group. 
Instead, we can exploit the would-be invariance  to choose the
arbitrary parameter $\mu$ in such a way as to maximally reduce its presence. Here,
it can obviously be done by taking $\mu =\gamma$ which makes the first term go away.
The resulting, total correction to the ground state energy is then
\beq
    \Delta E_0 &=& {\gamma^3\over \pi} \left({C_0^R\over 1 -  C_0^R I_\gamma^R}
               - {2\gamma^2C_2^R\over [1 - C_0^R I_\gamma^R]^2}\right) \nn \\
    &=& a m^2\alpha^3 \left[{2b\over b - a} - {a r_0\over (b - a)^2}\right] \label{dE2}
\eeq
where the first part is from (\ref{dE1}). 
It is possible also to obtain corrections to higher order in the effective range 
parameter. But we must then start out with an effective Lagrangian which involves 
higher derivative operators in the hadronic part (\ref{had}).

The obtained results (\ref{dE1}) and (\ref{dE2}) applies both to the spin singlet and
triplet component of the ground state since the total spin is conserved by the
strong  interactions. One can then obtain the hyperfine splitting in terms of the
two corresponding scattering lengths and effective ranges. Protonium is obviously very 
unstable and will decay via hadronic annihilation. As a consequence, the
energy levels will be smeared out because of these non-elastic channels. One can 
incorporate this effect in the present framework by letting the scattering lengths 
become complex\cite{Batty_1}. More phenomenological applications of the obtained results
to the protonium spectrum will be presented elsewhere together with higher order 
electromagnetic corrections\cite{KRnext}.

We are grateful to P. Labelle for stimulating advice on the applications of  
effective field theory to non-relativistic systems. Also H. Pilkuhn is acknowledged
for drawing our attention to hadronic atoms. We thank NORDITA for their generous
support and hospitality. Xinwei Kong is supported by the Research Council of Norway.

\end{document}